\documentclass[preprint]{aastex63}

\usepackage{color}

\newcommand\ISOIS{\hbox{IS\raise0.1ex\hbox{$\odot$}}IS}

\received{2019 September 27}
\revised{2019 October 31}
\accepted{2019 November 11}

\submitjournal{ApJS (for the {\it PSP} special issue)}

\shorttitle{2019 April 4 SEP Event at {\it PSP}}
\shortauthors{Leske et al.}

\begin{document}

\title{Observations of the 2019 April 4 Solar Energetic Particle Event at the {\it Parker Solar Probe}}

\author{R. A. Leske}
\affiliation{California Institute of Technology, Pasadena, CA 91125, USA}

\correspondingauthor{R. A. Leske}
\email{ral@srl.caltech.edu}

\author{E. R. Christian}
\affiliation{NASA Goddard Space Flight Center, Greenbelt, MD 20771, USA}

\author{C. M. S. Cohen}
\affiliation{California Institute of Technology, Pasadena, CA 91125, USA}

\author{A. C. Cummings}
\affiliation{California Institute of Technology, Pasadena, CA 91125, USA}

\author{A. J. Davis}
\affiliation{California Institute of Technology, Pasadena, CA 91125, USA}

\author{M. I. Desai}
\affiliation{Southwest Research Institute, San Antonio, TX 78228, USA}

\author{J. Giacalone}
\affiliation{University of Arizona, Tucson, AZ 85721, USA}

\author{M. E. Hill}
\affiliation{Johns Hopkins University Applied Physics Laboratory, Laurel, MD 20723, USA}

\author{C. J. Joyce}
\affiliation{Department of Astrophysical Sciences, Princeton University, Princeton, NJ 08540, USA}

\author{S. M. Krimigis}
\affiliation{Johns Hopkins University Applied Physics Laboratory, Laurel, MD 20723, USA}

\author{A. W. Labrador}
\affiliation{California Institute of Technology, Pasadena, CA 91125, USA}

\author{O. Malandraki}
\affiliation{National Observatory of Athens, IAASARS, Athens 15236, Greece}

\author{W. H. Matthaeus}
\affiliation{Department of Physics and Astronomy, University of Delaware, Newark, DE 19716, USA}
%\affiliation{University of Delaware, Newark, DE 19716, USA}

\author{D. J. McComas}
\affiliation{Department of Astrophysical Sciences, Princeton University, Princeton, NJ 08540, USA}

\author{R. L. McNutt Jr.}
\affiliation{Johns Hopkins University Applied Physics Laboratory, Laurel, MD 20723, USA}

\author{R. A. Mewaldt}
\affiliation{California Institute of Technology, Pasadena, CA 91125, USA}

\author{D. G. Mitchell}
\affiliation{Johns Hopkins University Applied Physics Laboratory, Laurel, MD 20723, USA}

\author{A. Posner}
\affiliation{NASA HQ, Washington, DC 20024, USA}

\author{J. S. Rankin}
\affiliation{Department of Astrophysical Sciences, Princeton University, Princeton, NJ 08540, USA}

\author{E. C. Roelof}
\affiliation{Johns Hopkins University Applied Physics Laboratory, Laurel, MD 20723, USA}

\author{N. A. Schwadron}
\affiliation{University of New Hampshire, Durham, NH 03824, USA}

\author{E. C. Stone}
\affiliation{California Institute of Technology, Pasadena, CA 91125, USA}

\author{J. R. Szalay}
\affiliation{Department of Astrophysical Sciences, Princeton University, Princeton, NJ 08540, USA}

\author{M. E. Wiedenbeck}
\affiliation{Jet Propulsion Laboratory,
California Institute of Technology, Pasadena, CA 91109, USA}

\author{A. Vourlidas}
\affiliation{Johns Hopkins University Applied Physics Laboratory, Laurel, MD 20723, USA}

\author{S. D. Bale}
\affiliation{Physics Department, University of California, Berkeley, CA 94720-7300, USA}
\affiliation{Space Sciences Laboratory, University of California, Berkeley, CA 94720-7450, USA}
\affiliation{The Blackett Laboratory, Imperial College London, London, SW7 2AZ, UK}
\affiliation{School of Physics and Astronomy, Queen Mary University of London, London E1 4NS, UK}
%\affiliation{University of California at Berkeley, Berkeley, CA 94720, USA}

\author{R. J. MacDowall}
\affiliation{Solar System Exploration Division, NASA/Goddard Space Flight Center, Greenbelt, MD 20771, USA}

\author{M. Pulupa}
\affiliation{Space Sciences Laboratory, University of California, Berkeley, CA 94720-7450, USA}

\author{J. C. Kasper}
\affiliation{Climate and Space Sciences and Engineering, University of Michigan, Ann Arbor, MI 48109, USA}
\affiliation{Smithsonian Astrophysical Observatory, Cambridge, MA 02138, USA}

\author{R. C. Allen}
\affiliation{Johns Hopkins University Applied Physics Laboratory, Laurel, MD 20723, USA}

\author{A. W. Case}
\affiliation{Smithsonian Astrophysical Observatory, Cambridge, MA 02138, USA}

\author{K. E. Korreck}
\affiliation{Smithsonian Astrophysical Observatory, Cambridge, MA 02138, USA}

\author{R. Livi}
\affiliation{University of California at Berkeley, Berkeley, CA 94720, USA}

\author{M. L. Stevens}
\affiliation{Smithsonian Astrophysical Observatory, Cambridge, MA 02138, USA}

\author{P. Whittlesey}
\affiliation{University of California at Berkeley, Berkeley, CA 94720, USA}

\author{B. Poduval}
\affiliation{University of New Hampshire, Durham, NH 03824, USA}

\begin{abstract}

A solar energetic particle event was detected by the Integrated Science
Investigation of the Sun (\ISOIS) instrument suite on
{\it Parker Solar Probe (PSP)} on 2019 April 4 when the spacecraft was inside of 0.17 au and less than 1~day before its second
perihelion, providing an opportunity to study solar particle acceleration and
transport unprecedentedly close to the source. The event was very small, with peak 1~MeV proton
intensities of $\sim$0.3 particles (cm$^2$ sr s MeV)$^{-1}$, and was undetectable above background
levels at energies above 10 MeV or in particle detectors at 1~au.  It was strongly anisotropic,
with intensities flowing outward from the Sun up to 30 times greater than those flowing inward
persisting throughout the event.
Temporal association between particle increases and small brightness surges in the extreme-ultraviolet 
observed by the {\it Solar TErrestrial RElations Observatory},
which were also accompanied by type III radio emission seen by the Electromagnetic Fields Investigation on {\it PSP}, indicates
that the source of this event was an active region nearly 80$^{\circ}$ east of the nominal {\it PSP} magnetic footpoint.
This suggests that the field lines expanded over a wide longitudinal range between the active region in the photosphere 
and the corona.

\end{abstract}


\section{Introduction} \label{sec:intro}

NASA's {\it Parker Solar Probe (PSP)} mission is designed to explore dynamic, energetic processes in the solar
corona and near-Sun interplanetary space \citep{fox}.  Launched in 2018 August into 
a highly elliptical solar orbit, it will utilize seven Venus gravitational assists over the course
of 7 yr to successively lower its perihelion, ultimately to under 10~R$_{\sun}$ (0.046~au).
Already by 2019 September it had made three perihelion passes at 0.166~au, closer than achieved
by any previous mission, and has observed several small solar energetic particle (SEP) events \citep{isoisnature}.
From its vantage point so much closer to the Sun, in addition to uncovering the origin of the
solar wind, it
is expected that its observations will help elucidate the mechanisms that accelerate SEPs
and govern their transport throughout the inner heliosphere, helping to disentangle the roles of
shocks, reconnection, waves, turbulence, and drifts in these processes.

 % ============================== FIGURE 1 ==============================
\begin{figure}[htb!]
\begin{center}
\includegraphics[trim=0 0 0 0,clip=true,width=0.5\textwidth]{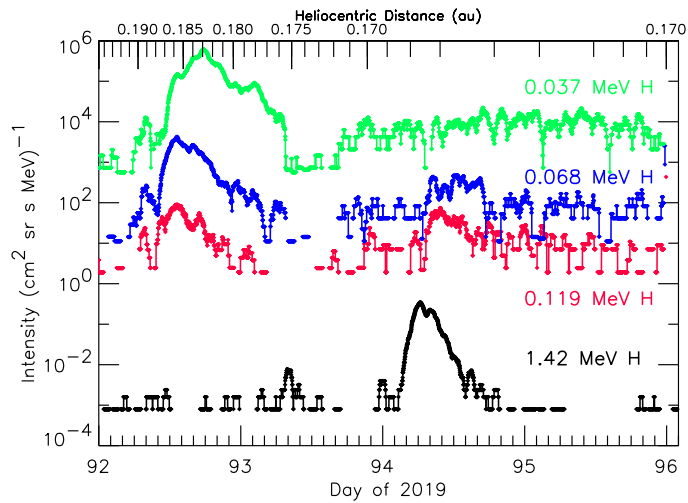}
\end{center}

\caption{Time profiles of proton intensities at the indicated energies 
for the period 2019 April 02 00:00--2019 April 06 02:00.
The top three traces are from {\ISOIS}/EPI-Lo time-of-flight measurements, 
while the bottom trace is from {\ISOIS}/EPI-Hi/LET1.  In all cases, 60-point moving averages
of minute-cadence data are shown.  The distance of {\it PSP} from the Sun during this
period is displayed on the top axis.\explain{``AU'' changed to ``au'' in top axis label.}}

\label{fig:overview}
\end{figure}
 % ====================================================================

During the solar encounter portion of {\it PSP}'s second orbit, a pair of small SEP events 
occurred 2 days apart (Figure~\ref{fig:overview}), originating from an active
region (AR) nearly 80$^{\circ}$ east of {\it PSP}'s nominal coronal magnetic footpoint.
The first, on April~2 (day of 2019 = 92 in the figure), was clearly detected by {\it PSP} instrumentation at energies below several hundred keV, 
but showed almost no increase in protons
at 1~MeV and above.  The second event, on April~4 (day 94), was observed when {\it PSP} was less than 0.17~au from the Sun and less than a day before
perihelion.  In contrast to the April~2 event, the increase in MeV protons was obvious, but it was much smaller in $<$100 keV particles and near background levels.
As is evident by the relative spacing of the traces in Figure~\ref{fig:overview} and the relative heights of the two events at a given energy, the April~4 event 
had a significantly harder proton energy spectrum than the April~2 event.  It is the April~4 event that is the focus of this paper; 
for more information on observations during the April~2 event and comparisons between the two, see \cite{roelof19}.

\section{Instrumentation} \label{sec:instrument}

The {\it PSP} Integrated Science Investigation of the Sun (\ISOIS) 
consists of a suite of two energetic particle instruments, together with their associated science 
operations, data processing, analysis tools, and personnel \citep{isois}.  One of the instruments, EPI-Lo, measures particles using the time-of-flight versus
energy technique and determines composition, spectra, and anisotropies of particles with energies from 
$\sim$20 keV nucleon$^{-1}$ to several MeV nucleon$^{-1}$ \citep{isois,mushroom}.  The other, EPI-Hi, is made up
of three silicon solid-state detector telescopes, two low-energy telescopes (LET1 and LET2), and one high-energy 
telescope (HET), that use the $dE/dx$ versus residual energy technique to
measure particles over the combined energy range of $\sim$1--200 MeV nucleon$^{-1}$ \citep{isois,wiedenbeck2017a}.
Each LET and HET telescope is cylindrical, with LET1 and HET being double-ended and LET2 single-ended.  Each of 
the five EPI-Hi apertures views a cone lying in the {\it PSP} orbital plane with a half-angle width of $\sim$45$^{\circ}$,
and each measures a subset of rates in up to 25 smaller directional sectors within its full field of view 
to provide anisotropy information \citep{isois}.  When the spacecraft is oriented with its long axis pointing at 
the Sun and turned so that EPI-Hi is on the leading side of the spacecraft (the orbital ram 
direction), as during solar encounters, one LET1 aperture (LET-A) is pointed 45$^{\circ}$ west of the Sun--spacecraft line (i.e., along
the nominal Parker spiral direction at 1~au) while the opposite end (LET-B) is pointed 135$^{\circ}$ east of the Sun.
LET2 is oriented orthogonal to LET1, that is, its single aperture normally points 135$^{\circ}$ west of the Sun.
One HET aperture points 20$^{\circ}$ west of the Sun with the opposite end 160$^{\circ}$ east of the Sun.
These fields of view for each EPI-Hi aperture are illustrated in Figure 18 of \cite{isois}.

Most of the particle data used in this study come from {\it PSP}/{\ISOIS}/EPI-Hi.  Although the April~4 event was observed
by EPI-Lo, it was near the instrument's sensitivity limit (see Figure~\ref{fig:overview}).  Detection efficiencies and backgrounds
in EPI-Lo are currently being assessed \citep{hill19} and intercalibration with EPI-Hi is underway \citep{joyce19}; preliminary
EPI-Lo observations during this event are presented elsewhere \citep{isoisnature,roelof19,joyce19}.
As we show in Section~\ref{sec:obs:parts}, the 2019 April~4 event studied here was very small, anisotropic, and 
spectrally soft.  With a proton energy threshold of $\sim$8~MeV, EPI-Hi/HET did not detect the event at all.  As
mentioned above, the LET2 telescope points away from the Sun (in a direction where particle intensities were lower), and it also
has a thicker entrance window (and hence a higher energy threshold) than LET1; as a result, the measured count rates 
in LET2 were more than an order of magnitude lower than in LET1 during this event.  Therefore, all of the
EPI-Hi data shown here come exclusively from LET1, both A and B apertures. 

% LET1 proton threshold: ~1.15 MeV
% LET2 proton threshold: ~1.8 MeV, plus pointed away from peak direction --> <1/10 intensity of LET1A
% HET proton threshold: ~8 MeV --> event not seen at all

\section{Observations} \label{sec:obs}

\subsection{In Situ Observations} \label{sec:obs:parts}

\begin{figure}[htb!]
\begin{center}
\plottwo{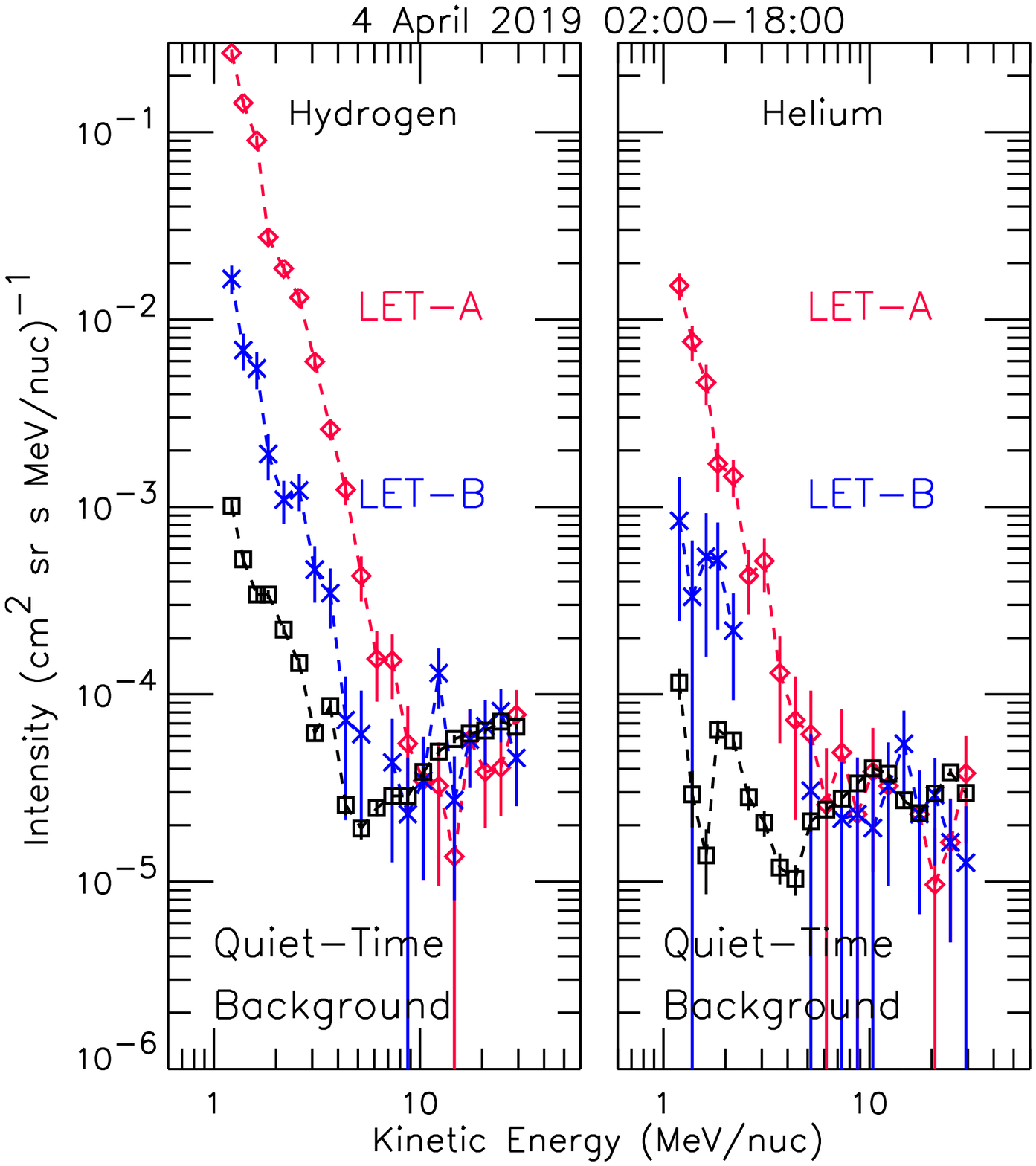}{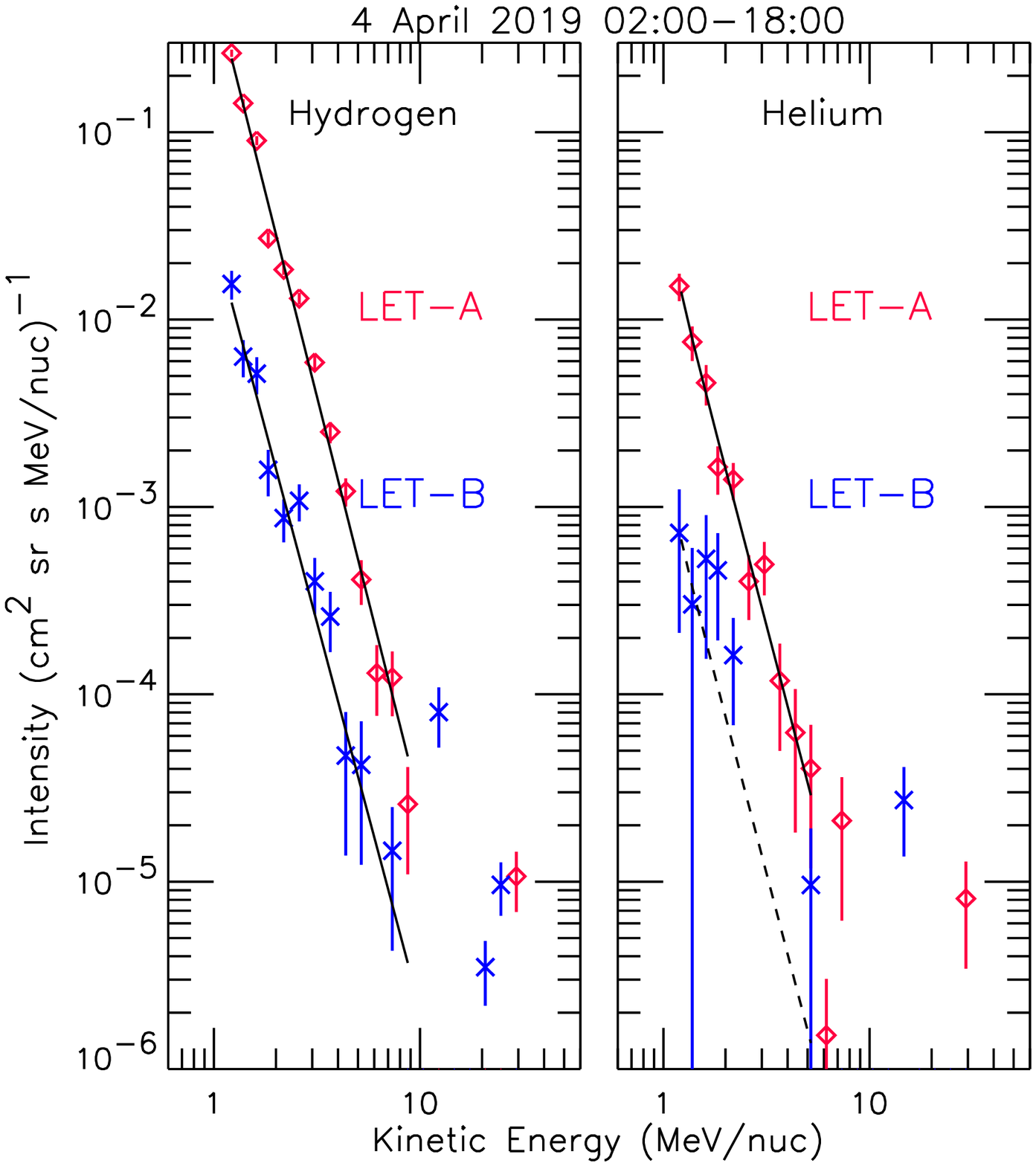}
\end{center}

\caption{Hydrogen and helium differential energy spectra averaged over the duration of
the 2019 April~4 SEP event (02:00--18:00) from {\ISOIS}/EPI-Hi/LET1.  Spectra measured in the
aperture facing generally toward the Sun (LET-A) are in red, while those in the opposite aperture (LET-B) 
are in blue; a large antisunward anisotropy results in much higher intensities in LET-A than in LET-B.  
Backgrounds due to quiet-time particles unrelated to the SEP event (see text) are shown in black 
in the left pair of panels and are subtracted from the SEP data in the right pair of panels.  Power-law fits
are shown in the right panels as solid black lines \added{over the energy interval used in the fit}; the dashed line for LET-B helium is simply the LET-A helium
fit scaled downward by the fit hydrogen LET-A/LET-B ratio, a factor of 21.}

\label{fig:spectra}
\end{figure}

The hydrogen and helium elemental energy spectra from EPI-Hi, averaged over the 2019 April~4 SEP event from 02:00 to 18:00,
are compared in Figure~\ref{fig:spectra} for LET-A and LET-B.  To estimate the quiet-time background levels, a loose selection of quiet periods
was made by requiring that the daily averaged proton intensity near 3~MeV (specifically, in the 2.83--3.36~MeV energy
bin) be $<$8.9$\times$10$^{-5}$ particles (cm$^{2}$ sr s MeV)$^{-1}$.  This cut excludes SEP events
(such as those discussed in \cite{isoisnature,giacalone19,wiedenbeck19,schwadron19}) as well as intensity enhancements due to particles associated with
stream interaction regions (SIRs) and corotating interaction regions (CIRs) \citep{cohen19}.  The resulting ``quiet'' intervals total to
$\sim$34~days, and the H and He spectra averaged over this period
are shown in black in the
left pair of panels in Figure~\ref{fig:spectra} (LET-A and LET-B quiet spectra were examined separately and found to be consistent, 
and so they have been averaged together).  Above $\sim$5~MeV, the quiet-time proton spectrum has both the intensity and spectral
shape expected for galactic cosmic rays (GCRs).  The turn-up at lower energies is not completely understood, but it may
contain contributions from misidentified, out-of-geometry high-energy GCRs or smaller SIR or CIR events that were missed by the
loose quiet cut; in any case this unidentified background is only $\sim$1\% of the LET-A and $\sim$10--20\% of the LET-B proton intensity during the April~4 event. 
Similarly, the intensity of the quiet-time He spectrum above 5~MeV nucleon$^{-1}$ is about as expected from the combination of anomalous cosmic ray (ACR) He 
and GCR He, but the peculiar wiggle in the shape of the spectrum indicates some sort of instrumental effect, which is under investigation but of no concern for this 
study since the SEP spectrum in this event does not extend to these energies.  The low-energy He quiet-time spectrum suggests the presence of 
unidentified backgrounds or instrumental effects, but it is at $<$10\% of the LET-A He SEP spectrum and thus relatively unimportant.  

 % ============================== FIGURE 2 ==============================
 % ====================================================================

In the pair of panels on the right side of Figure~\ref{fig:spectra}, the background-corrected spectra
are shown.  At this early stage of the mission during this quiet solar minimum, with very limited high-energy particle data 
with which to evaluate instrument performance, instrumental effects are still present
in these spectra (most notably the dip around 2~MeV nucleon$^{-1}$).  Nevertheless, the spectra
seem consistent with a single power law over the energy range of $\sim$1--8~MeV nucleon$^{-1}$.  Within formal statistical
uncertainties, the spectral indices for the power-laws fits for LET-A H, LET-B H, and LET-A He are the same, 
at $-4.36\pm0.06$, $-4.12\pm0.26$, and $-4.21\pm0.26$, respectively.  Statistics for LET-B He are too meager to 
permit a sensible fit.  For illustration purposes, the LET-A He fit has been scaled by the 1~MeV H LET-B-to-LET-A ratio.
The result suggests that the spectral shape of LET-B He may be similar to that of LET-A He, and that the B-to-A He ratio
is similar to that for H.

Preliminary intercalibration of EPI-Hi and EPI-Lo \citep{joyce19} indicates that the proton power-law spectra shown
here do not continue down to 100~keV with the same index.  Instead, there seems to be a spectral break or rollover
somewhere between $\sim$300~keV and 1~MeV, with a harder spectral index below.  \added{At present, the EPI-Lo proton
efficiency and background are too uncertain to justify attempting to fit a single functional form to the
combined spectra from EPI-Lo and EPI-Hi in this very small event.  Furthermore, the extreme anisotropy, along
with the fact that the fields of view of the two instruments are different \citep{isois}, would make such a 
fit problematic.} 
\deleted{Further details of the EPI-Lo measurements in this event are available in \cite{roelof19}.}

\begin{figure}[htb!]
\begin{center}
\includegraphics[trim=0 0 0 0,clip=true,width=0.3\textwidth]{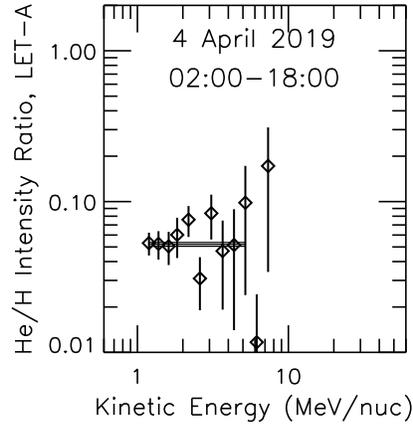}
\end{center}

\caption{He/H ratio measured in {\ISOIS}/EPI-Hi/LET-A as a function of energy, after quiet-time background
subtraction, averaged over the duration of the 2019 April~4 SEP event.
The average value of 0.052$\pm$0.002 is marked by the horizontal lines.}

\label{fig:heh}
\end{figure}

After the background subtraction in Figure~\ref{fig:spectra}, the He/H ratio is found to be 0.052$\pm$0.002, as shown in Figure~\ref{fig:heh}.  
This value is 
well determined at 1--2~MeV nucleon$^{-1}$, and consistent with being independent of energy, but the low statistics and resulting large
uncertainties mean that a sizable increase or decrease in this ratio at energies from 2 to 8~MeV nucleon$^{-1}$ cannot be ruled out.

 % ============================== FIGURE 3 ==============================
 % ====================================================================

 % ============================== FIGURE 4 ==============================
\begin{figure}[htb!]
\begin{center}
\includegraphics[trim=0 0 0 0,clip=true,width=0.3\textwidth]{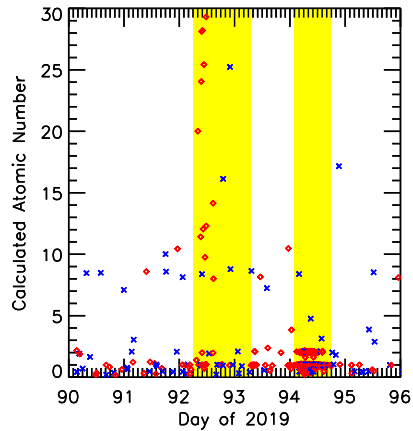}
\end{center}

\caption{Calculated atomic number vs.\ arrival time for particles entering LET-A (red diamonds) and LET-B (blue crosses)
with energies of $\sim$1--5 MeV nucleon$^{-1}$, using pulse-height data.  Yellow bands indicate the time periods
of the April~2 (left) and April~4 (right) SEP events.  The calibration is preliminary, so the calculated atomic
number is only approximate, particularly for the heavier ions.}

\label{fig:zvst}
\end{figure}
 % ====================================================================

Although EPI-Hi is capable of measuring electrons, protons, and heavier elements up through nickel, as well as 
He isotopes \citep{isois}, no electrons above 0.5 MeV or ions heavier than $^{4}$He were detected
above quiet-time background levels, and no $^{3}$He was observed in this very small SEP event.  Using the
same analysis employed by \cite{wiedenbeck19} for their study that made a definitive $^{3}$He detection
in the much larger 2019 April~21 SEP event, we find 0~$^{3}$He nuclei and 19~$^{4}$He nuclei in the narrow energy range
of 1.35--1.75 MeV nucleon$^{-1}$ at near-normal incidence angles.  Although this results in a formal 1$\sigma$
upper limit (using Poisson statistics) of 0.097 for the $^{3}$He/$^{4}$He ratio, we find no strong evidence of additional $^{3}$He at 
higher energies, suggesting that with further analysis a lower limit could be obtained.

The absence of ions heavier than He in this event is interesting, since the He intensity
reached two orders of magnitude above background (and background levels appear to be lower for the heavier ions), 
which would have been large enough to detect some heavies if their
abundances relative to He were typical of those seen in SEPs \citep{reames14}.
As shown in Figure~\ref{fig:zvst}, during the entire period of 2019 April~4 02:00--18:00, there were no LET-A heavy ions detected below 
5 MeV nucleon$^{-1}$, one oxygen nucleus in LET-B (consistent with the expected quiet-time ACR oxygen background), and two LET-B particles 
with an apparent atomic number between 2 and 5, probably due to instrumental background of the sort suggested to explain the quiet-time low-energy
proton increase discussed above.  Based on the 0 detected counts, the 1$\sigma$ Poisson upper limit for the 
LET-A 1--5 MeV nucleon$^{-1}$ oxygen intensity during this period is $3.2\times10^{-5}$ particles (cm$^{2}$ sr s MeV nucleon$^{-1}$)$^{-1}$.
Since the He intensity at the mean energy in this interval is $\sim5\times10^{-3}$ particles (cm$^{2}$ sr s MeV nucleon$^{-1}$)$^{-1}$ (Figure~\ref{fig:spectra}),
the O/He ratio must be $<0.006$ (and after accounting for expected ACR oxygen background, this would drop to $\sim$0.004).
This value is significantly lower than the average value reported
in SEPs of $\sim$0.02 \citep{reames14}, and closer to spectroscopically measured coronal and photospheric values of 8$\times$10$^{-3}$ \citep{feldman03} 
and 6$\times$10$^{-3}$ \citep{caffau11,lodders10}, respectively.  In contrast, during the April~2 event when EPI-Hi saw almost no increase in protons (Figure~\ref{fig:overview}) and 
the 1~MeV nucleon$^{-1}$ He intensity was $\sim$30 times lower, LET-A detected 12 heavy nuclei up to Fe below 5~MeV nucleon$^{-1}$, with another 2 nuclei heavier than oxygen
in LET-B (Figure~\ref{fig:zvst}).
Thus, not only was the O/He ratio at MeV energies in the April~4 event somewhat lower than typical SEP values, but the heavy-ion-to-He ratio was also at least $\sim$300 times lower 
than in the event only two days earlier from the same AR.  See \cite{roelof19} for further discussion of the
characteristics of the composition in the April~2 event and comparison with the April~4 event, particularly at lower (EPI-Lo) energies.

 % ============================== FIGURE 5 ==============================
\begin{figure}[htb!]
\begin{center}
\includegraphics[trim=0 0 0 0,clip=true,width=0.5\textwidth]{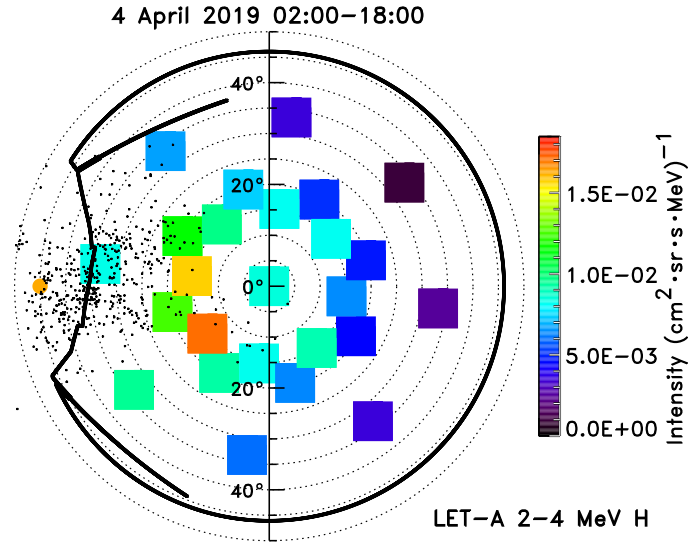}
\end{center}

\caption{Directional intensities of 2-4~MeV protons averaged over the 2019 April~4 SEP event in the 25 LET-A sectored viewing directions
are shown using the color scale on the right.  On this polar plot, values of the radial coordinates indicate the angle (marked on
the axis) from the telescope boresight, while angular coordinates correspond to the azimuthal angle about the telescope axis, with the vertical direction normal to the spacecraft orbit.
The colored squares are shown centered on the nominal sector viewing directions, but their shapes and sizes are not representative of those of the
sectors themselves, which overlap considerably \citep{isois} and together fill the region outlined by the thick black curve, which marks the 
boundary of the full LET-A field of view.  (Along the left-hand side, the spacecraft thermal protection system and FIELDS antennas obstruct part of the field of view).
Black dots mark the direction of the magnetic field as measured by FIELDS, using 1-minute data over the period 04:00--12:00, when most of the particles 
appear (see Figure~\ref{fig:timeprofiles}).  The filled orange circle at the left indicates the position of the Sun on the same angular scale.}

\label{fig:anisomap}
\end{figure}
 % ====================================================================

%IDL> print,max(rr2a/gflet1ar2.ogf)/min(rr2a/gflet1ar2.ogf)
%      16.9736

\begin{figure}[htb!]
\begin{center}
\includegraphics[trim=0 0 0 0,clip=true,width=0.5\textwidth]{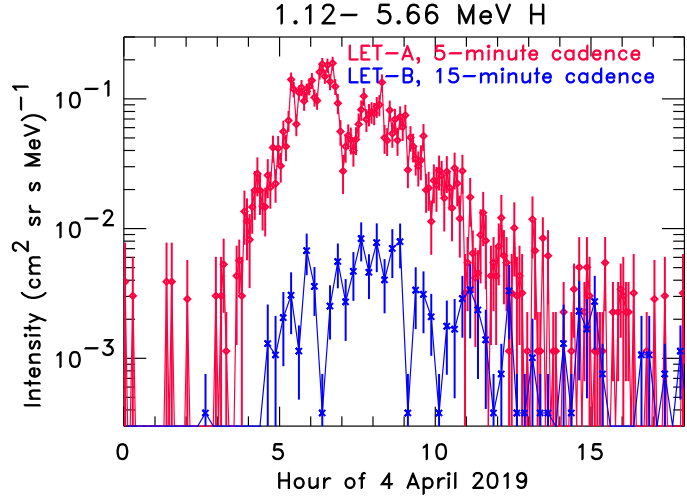}
\end{center}

\caption{Time profiles of 1.12--5.66~MeV hydrogen from {\ISOIS}/EPI-Hi/LET1 at a 
5-minute cadence for LET-A (red) and 15-minute cadence for LET-B (blue)
during the 2019 April~4 SEP event.  Statistical uncertainties are shown on each point.}

\label{fig:timeprofiles}
\end{figure}

The SEP event is highly anisotropic, with event-averaged intensities in Figure~\ref{fig:spectra} coming from the Sun in LET-A $\sim$21 times
higher than those heading toward the Sun in LET-B (early in the event the anisotropy was even greater, as 
presented later in Figure~\ref{fig:timeprofiles}).  Furthermore, the intensities are not uniform throughout the LET-A aperture, as shown in Figure~\ref{fig:anisomap},
but instead vary relatively smoothly by more than an order of magnitude across its field of view.  We have not yet determined pitch-angle distributions;
at this early stage of the analysis the sectored geometry factors used were calculated with a preliminary map of field-of-view obstructions for an isotropic distribution
and not yet iterated based on the measured anisotropy, deconvolved to disentangle the overlapping fields of view, or had Compton--Getting corrections
applied to convert to the solar wind rest frame (as discussed in \cite{roelof19}), so the intensity
distribution in Figure~\ref{fig:anisomap} is not exact.  Nevertheless, the intensities do seem to peak somewhere near the magnetic field
directions measured by the Electromagnetic Fields Investigation (FIELDS; \cite{fields}) magnetometer on {\it PSP}, as would be expected for a field-aligned beam of particles.

 % ============================== FIGURE 6 ==============================
 % ====================================================================

Time profiles of 1.12--5.66~MeV protons on both LET-A and LET-B are shown in Figure~\ref{fig:timeprofiles}.
The entire event is very brief, lasting only $\sim$15~hr, and takes $\sim$4~hr from onset to reach peak intensities.
The most notable structure in the event is a pronounced dip in LET-A intensities by a factor of $\sim$5 around 07:00.  Near the
peak of the event, the LET-A intensities are a factor of $\sim$30 times greater than those in LET-B.  
Although particles are not detected in LET-B until $\sim$1.5 hr after they first arrive in LET-A, the very limited
statistics in LET-B make it difficult to determine whether this is a genuine onset delay; scaling from the LET-A intensities
suggests that only $\sim$2 additional particles in LET-B between $\sim$3:45 and 4:30 would give the same onset in both apertures.
Similarly, it is unclear whether or not the dip at 07:00 in LET-A is completely absent from LET-B.

 % ============================== FIGURE 7 ==============================
\begin{figure}[htb!]
\begin{center}
\includegraphics[trim=0 0 0 0,clip=true,width=0.5\textwidth]{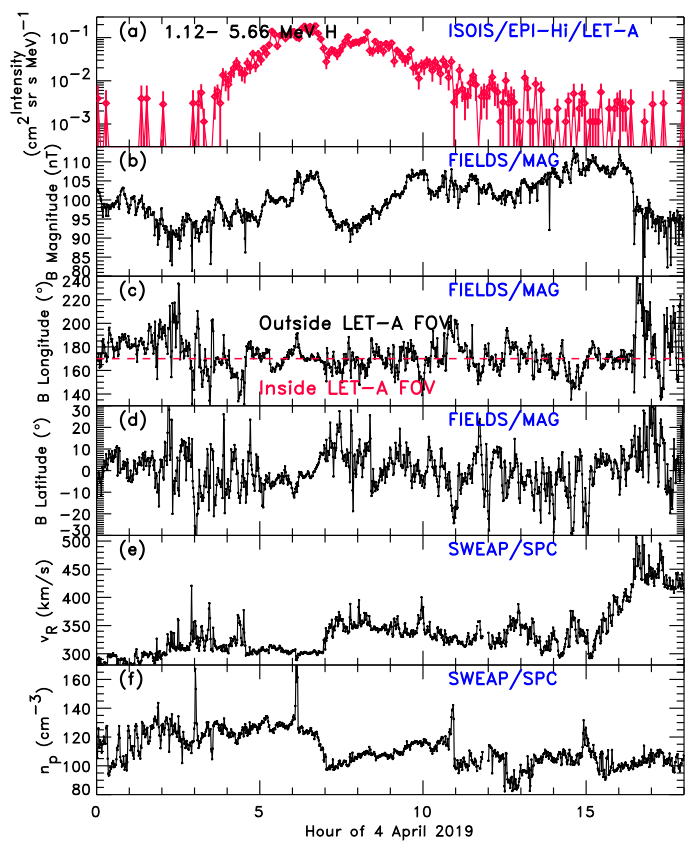}
\end{center}

\caption{Time profile of 1.12--5.66~MeV hydrogen from {\ISOIS}/EPI-Hi/LET-A at a 
5-minute cadence (as in Figure~\ref{fig:timeprofiles}) in panel (a) is compared with the
1-minute cadence magnetic field magnitude (panel (b)), RTN magnetic longitude (panel (c)), and RTN
magnetic latitude (panel (d)) from the {\it PSP}/FIELDS magnetometer, and with the 1-minute cadence
radial solar wind speed (panel (e)) and solar wind proton density (panel (f)) from the {\it PSP}/SWEAP/SPC
instrument.  The approximate demarcation between longitudinal directions inside and outside
the LET-A field of view is indicated by the dashed red line in panel (c).}

\label{fig:timeprofilesmagsw}
\end{figure}
 % ====================================================================

 % ============================== FIGURE 8 ==============================
\begin{figure}[htb!]
\begin{center}
\includegraphics[trim=0 0 0 0,clip=true,width=0.5\textwidth]{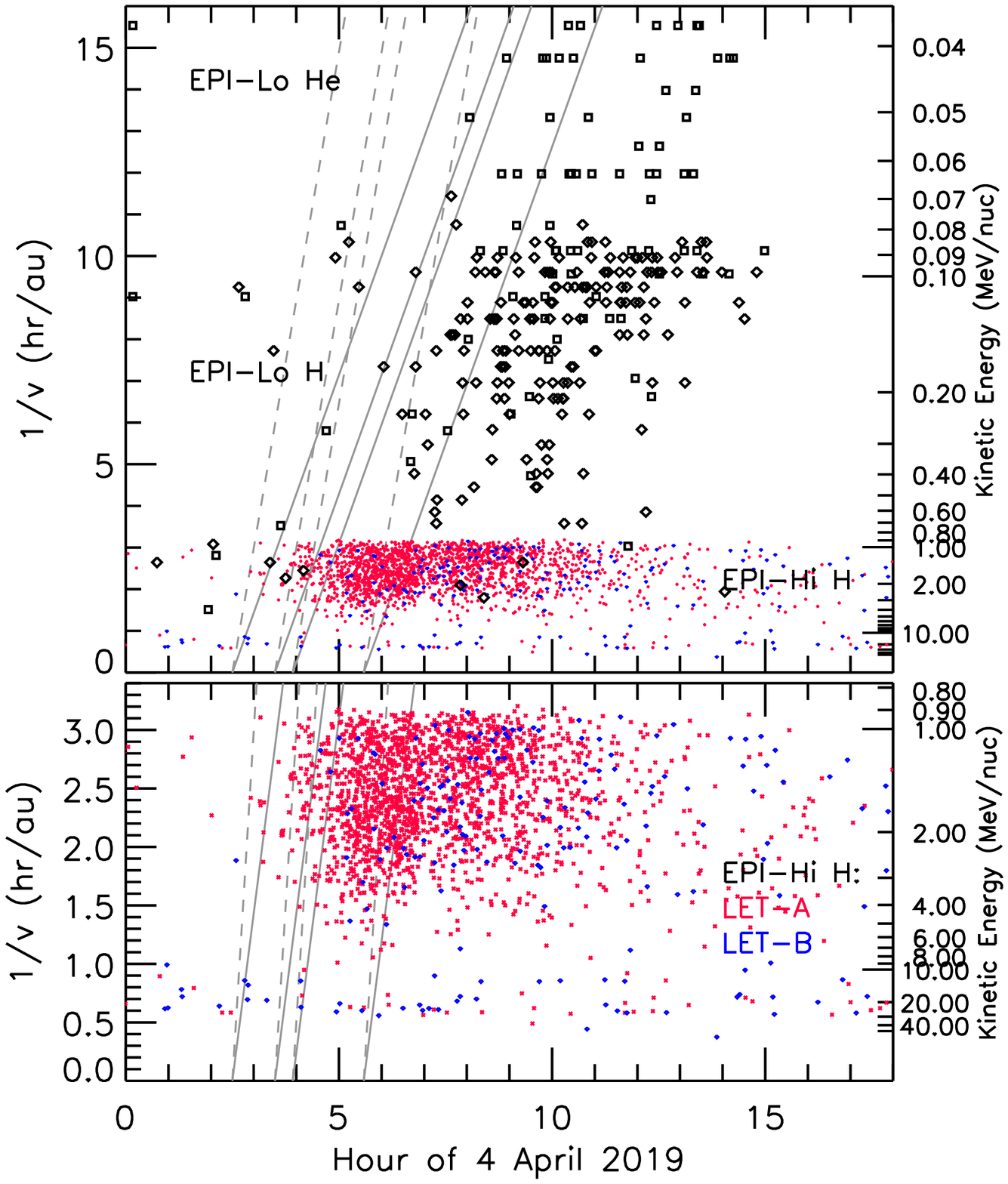}
\end{center}

\caption{Top panel: The reciprocal of the particle velocity plotted versus detection time for {\ISOIS}/EPI-Lo
1-minute helium rates (black squares) and hydrogen rates (black diamonds), and for {\ISOIS}/EPI-Hi
proton pulse-height event data from LET-A (red) and LET-B (blue).  The energy corresponding to the
reciprocal velocity is shown on the right axis.  A symbol is plotted for EPI-Lo when the
corresponding rate is nonzero; no distinction is made for counts greater than 1.
Tilted gray lines indicate the expected location of particles
if they left the Sun at the time of any of the four {\it STEREO}/EUVI surges identified in Figure~\ref{fig:euvi} and traveled
without scattering for a distance equal to that of the nominal Parker spiral at perihelion for a solar wind speed 
of 350 km s$^{-1}$, 0.168 au (dashed lines), or a much longer
path length of 0.35 au (solid lines).  Bottom panel: An enlarged view of only the EPI-Hi data from the top panel.
\explain{``hr/AU'' changed to ``hr/au'' in vertical axis labels.}}

\label{fig:vdisp}
\end{figure}
 % ====================================================================

The SEP time profile from Figure~\ref{fig:timeprofiles} is compared with variations in solar wind parameters in
Figure~\ref{fig:timeprofilesmagsw}, using data from the FIELDS magnetometer \citep{fields} and from the
Solar Proton Cup (SPC) of the Solar Wind Electrons Alphas and Protons suite (SWEAP; \cite{sweap}) on {\it PSP}.
As shown in panel (c), roughly half the time during this event the magnetic field longitude is outside the
LET-A field of view, by up to $\sim$30$^{\circ}$ (this is also seen in Figure~\ref{fig:anisomap}).  If the proton pitch-angle 
distribution were aligned with 
the field and narrow compared to the 45$^{\circ}$ half-angle LET-A aperture, movement of the beam out of the
LET-A field of view would result in a decrease 
in the measured particle intensity, but the observed dip in protons near 07:00 does not seem to be associated with
any obvious change in the average field longitude.  (In contrast, the large change in field longitude at $\sim$16:30 essentially marks
the end of the particle event in EPI-Hi).  The dip is, however, temporally correlated with small changes in the
solar wind: a 15\% decrease in magnetic field magnitude, 40\% decrease in solar wind density, and 15\% (50 km s$^{-1}$) increase in solar wind
speed as the spacecraft apparently entered a very small, weak rarefaction region\footnote{At this point near perihelion the spacecraft was moving faster 
than corotation, unlike the more familiar case when an observer changes heliolongitude slowly compared with the
solar rotation rate.  Since {\it PSP}'s Carrington longitude was moving westward, it would overtake a rarefaction region by approaching
it from the eastern, slow-speed side, and the solar wind speed would increase upon entering the rarefaction.} or encountered a small transient.

Note that in Figure~\ref{fig:overview} the higher-energy protons in the April~4 event start to 
increase hours before the lower-energy protons.  This velocity dispersion is examined more closely
in Figure~\ref{fig:vdisp}, where the reciprocal of the particle velocity is plotted versus the 
arrival time for EPI-Lo He and H rates and for EPI-Hi proton pulse-height data.  If the particle
injection profile at the source were a step function and they traveled without scattering, the
first-arriving particles at each energy would form a sharp, linear edge on such a plot.  The slope 
of that edge would give the path length the particles traveled, while the intercept on the horizontal
axis would indicate the particle release time at the Sun.  Although the combined data show some indication of 
an onset slope, the edge is very indistinct, perhaps in part due to low statistics.  For comparison,
dashed lines show where such an onset edge would be if the particles left the Sun at the time of the observed solar activity described
in Section~\ref{sec:obs:euvi} and traveled scatter-free along a magnetic field line with a length equal to that 
of the nominal Parker spiral from the corona to {\it PSP} (0.168 au).  Solid lines corresponding to a much larger path length of 0.35 au are also shown,
which seem to come closer to reproducing the observed slopes.  Our interpretation of these observations is discussed in Section~\ref{sec:disc}.

 % ============================== FIGURE 8 ==============================
\begin{figure}[htb!]
\begin{center}
\includegraphics[trim=0 0 0 0,clip=true,width=0.35\textwidth]{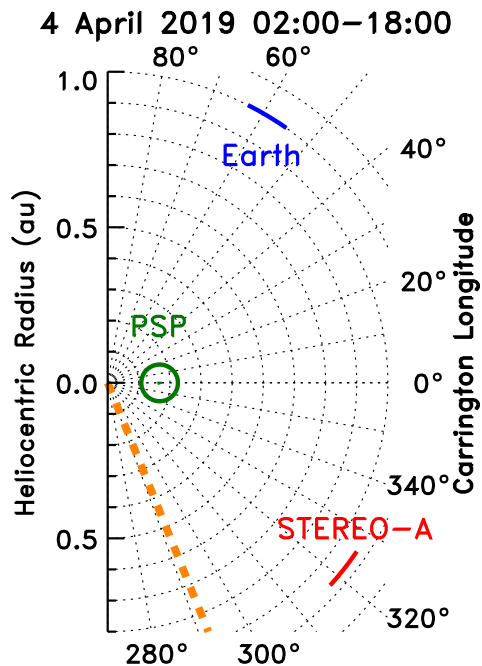}
\end{center}

\caption{Locations of Earth (blue), {\it STEREO-A} (red) and {\it PSP} (green; circled) in
heliographic radius and Carrington longitude during the period 02:00--18:00
on 2019 April~4.  During this interval, both Earth and {\it STEREO-A} move $\sim$8.2$^{\circ}$
clockwise on this plot (eastward in longitude), while {\it PSP} moves $\sim$2.9$^{\circ}$
counterclockwise (westward).  The longitude of AR 12738 is indicated by
the dashed orange line.\explain{``AU'' changed to ``au'' in vertical axis label.}}

\label{fig:location}
\end{figure}
 % ====================================================================

\subsection{Solar Observations} \label{sec:obs:euvi}

\begin{figure}[htb!]
\begin{interactive}{animation}{2190404_comb.mov}
\begin{center}
\includegraphics[trim=0 0 0 0,clip=true,width=0.55\textwidth]{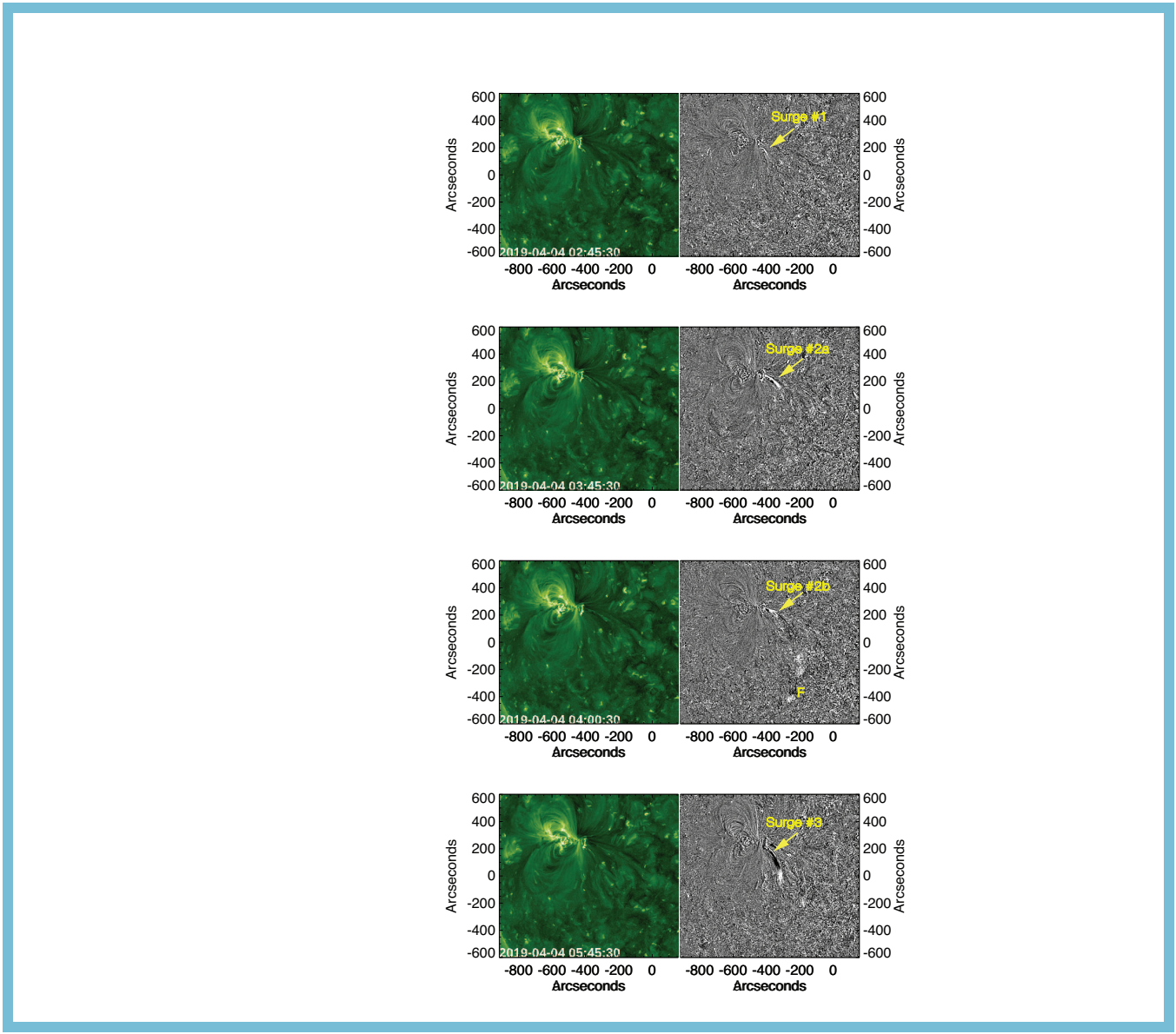}
%\fbox{\includegraphics[trim=72 100 80 0,clip=true,width=0.98\textwidth]{euvifig.pdf}}
%\plotone{euvifig.pdf}
\end{center}
\end{interactive}

\caption{{\it STEREO-A} EUVI 195\AA\, observations of the solar activity related to the April~4 SEP event in direct (left
panels) and running difference (\added{5 minutes apart,} right panels) images. The EUVI images have been enhanced by a wavelet algorithm to remove the 
stray-light background \citep{stenborg08}.  Snapshots of four surges at their times of maximum visibility are shown here; the first appearances
in EUVI for surges \#1, \#2a, \#2b, and \#3 are at about 02:30, 03:30, 03:55, and 05:35, respectively.  A 9 s video that 
includes these frames and covers 6 hr of observations beginning at
01:00 on 2019 April~4 is available in the online version of the paper.\explain{Horizontal and vertical scales added to figure.}}

\label{fig:euvi}
\end{figure}

Due to the location of {\it PSP} at the time, observations from the Extreme UltraViolet Imager (EUVI; \cite{secchi}) on the Ahead {\it Solar TErrestrial RElations
Observatory} ({\it STEREO-A}; \cite{stereo}) spacecraft provide the only means to locate the source of the event
detected by EPI-Hi on 2019 April~4.
At 03:00 on this date, near the particle event onset, 
{\it STEREO-A} was at 324$^{\circ}$ Carrington longitude and {\it PSP} at 359$^{\circ}$ (as shown in Figure~\ref{fig:location}), only
about 35$^{\circ}$ apart in longitude and 1$^{\circ}$ apart in latitude, so most of the solar hemisphere facing {\it PSP} was within
{\it STEREO}'s field of view.  The very quiet
solar conditions make the identification of the solar origins quite straightforward.  There was only one
AR visible to {\it STEREO-A} on the disk, located at N06, $L=292$ (in Carrington coordinates), 32$^{\circ}$ east of
{\it STEREO}'s central meridian, and numbered AR 12738 when it rotated into view of
Earth a few days later.  This same region was the source of the event detected by EPI-Lo on April~2 \citep{roelof19} and of much 
larger SEP events observed by {\ISOIS} and at 1~au on April~20--21 \citep{wiedenbeck19,schwadron19}.  The AR remained active
throughout the {\it PSP} perihelion passage, with multiple narrow ejections occurring daily.  We focus
on the activity during the first 7 hr of April~4 that is most relevant to the in situ
observations discussed here.  The ejections were very similar to each other.  They were narrow,
were curved, and contained cold material seen in absorption.  This morphology is characteristic of
H$\alpha$ surges (e.g., \cite{roy73,schmahl81,chae99}) and we adopt this interpretation in this paper.  All surges originated from an area on the
AR western edge and were directed toward the southwest (Figure~\ref{fig:euvi}), roughly toward the nominal {\it PSP} coronal magnetic footpoint.

 % ============================== FIGURE 10 ==============================
 % ====================================================================

\begin{figure}[htb!]
\begin{center}
\includegraphics[trim=0 0 0 0,clip=true,width=0.8\textwidth]{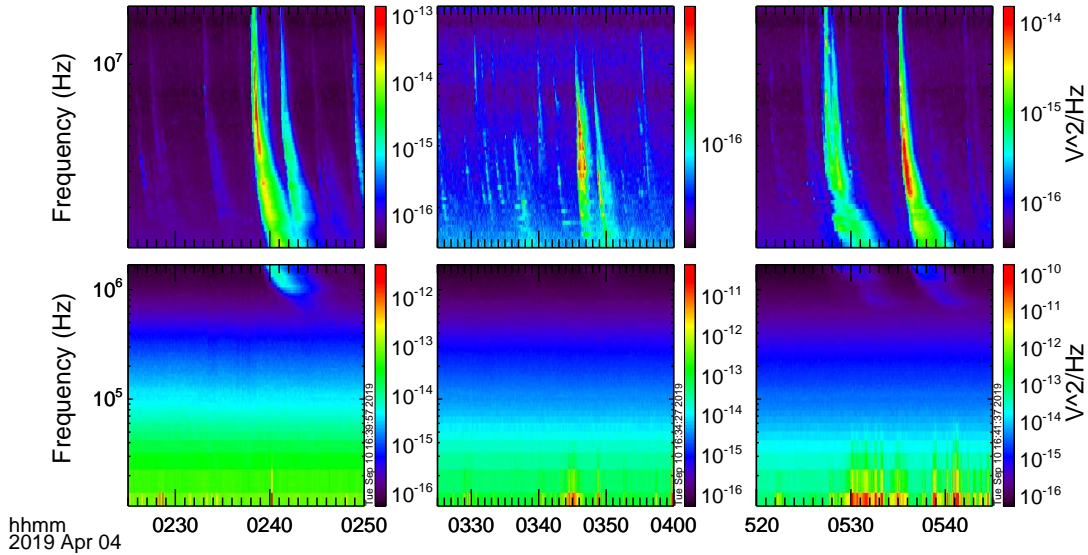}
\end{center}

\caption{Dynamic radio spectra from the {\it PSP}/FIELDS/RFS at high frequencies (top) and low frequencies (bottom) 
around the times of the {\it STEREO}/EUVI surges shown in Figure~\ref{fig:euvi}.  Type III signatures corresponding to surge \#1
are shown in the left panels, surges \#2a and \#2b in the middle panels, and surge \#3 in the right panels.
Note the different time and intensity scales in each panel; the radio emission during surges \#2a and \#2b
is significantly weaker than for the others.\explain{Vertical axis labels simplified.}}

\label{fig:radio}
\end{figure}

We identified four surge events between 02:00 and 07:00 UT, as labeled in Figure~\ref{fig:euvi}.
All have counterpart type III signatures in the FIELDS/Radio Frequency Spectrometer (RFS; \cite{fields,pulupa}) \explain{Pulupa et al. reference added} spectra, shown in Figure~\ref{fig:radio}.  
Note that the times reported for EUVI have a 5-minute cadence, while the RFS spectra used here have subminute resolution.  Although they have
unmistakable type III signatures, surges \#1 
%(appearing in EUVI at $\sim$02:30 UT, with a pair of type III bursts seen by FIELDS at 02:38 and 02:41) 
and \#3 
%(appearing at $\sim$05:35 UT, with type III bursts at 05:27 and 05:35), 
are unremarkable in terms of material ejection. 
The most massive are surges \#2a and \#2b,
%(ejected at 03:30 and 03:55, respectively, with weak type III emission at 03:45 and 03:48), 
which occur in succession and exhibit the clearest
signatures of plasma motion.  The plasma moves along a curved southwestern trajectory and
appears to return to the surface at point F (see online movie accompanying Figure~\ref{fig:euvi}).  No waves can
be detected for any of these surges.
No coronal mass ejections (CMEs) are detected from either {\it STEREO-A} or the Large Angle and Spectrometric Coronagraph (LASCO; \cite{lasco}) 
on board the {\it Solar and Heliospheric Observatory} \citep{soho} satellite (only LASCO/C3 is available during this
period owing to a campaign). The lack of CMEs is consistent with the lack of surface
manifestations, such as waves or dimmings.  Large-enough surges are usually seen as jets (e.g.
\cite{vourlidas03}), which are narrow. 
The LASCO/C3 images show faint, narrow outflows between 06:30 and 11:10 UT along
approximately the {\it PSP} position angle, but it is difficult to say whether they are associated with the
radio or in situ signatures discussed here.  

\added{Previous studies have reported ``mini-CMEs'' and other
associated very small-scale activity such as mini-waves and mini-dimmings at a
high rate of $\sim$1 event per minute distributed across the quiet Sun (e.g., \cite{innes,pod2010}).  
While we cannot rule out their presence, such activity has not been associated with the production 
of escaping particles accelerated to MeV energies.  In contrast, the surges we identify here are
the largest manifestations of solar activity during this period, originate from the only AR on the
disk, and are associated with type III emission.  As shown in Figure~\ref{fig:vdisp}, they are temporally
associated with the April~4 SEP event.}

 % ============================== FIGURE 11 ==============================
 % ====================================================================
%++++++++++++++++++++++++++++++
\section{Discussion} \label{sec:disc}

Peak MeV proton intensities in the very small 2019 April~4 SEP event are similar to those in the many SIR events seen
during the {\it PSP} mission so far \citep{cohen19}. 
Although the relatively soft proton spectral index of -4.36 is somewhat harder than the average SIR value of $\sim$-5 in the \cite{cohen19}
study, what unmistakably marks this as an SEP event is the large antisunward anisotropy, significant
velocity dispersion, and associated solar activity (EUVI surges and type III radio emission).

Studies based on previous multispacecraft observations find the peak intensity in an SEP event to 
decrease with heliocentric radius, $R$, as $R^{-\alpha}$, where typically $\alpha<3$ \citep{lario13rad}.  If a spacecraft
at 1~au were magnetically well connected to {\it PSP}, where peak MeV proton intensities were $\sim$0.2 protons (cm$^2$ sr s MeV)$^{-1}$
(Figure~\ref{fig:timeprofiles}), one might therefore expect it to have seen an event with 
a peak intensity of at least $\sim$10$^{-3}$ protons (cm$^2$ sr s MeV)$^{-1}$.  The Low Energy Telescope on {\it STEREO-A} \citep{stereolet}
saw no increase in MeV protons above its background level of $\sim$5$\times10^{-5}$ protons (cm$^2$ sr s MeV)$^{-1}$, but its nominal magnetic footpoint in the corona
(for the observed solar wind speed of 350 km s$^{-1}$)
 was $\sim$23$^{\circ}$ west of that of {\it PSP}, or over 100$^{\circ}$ west of the AR (near-Earth spacecraft were even farther away in longitude
than {\it STEREO}, as shown in Figure~\ref{fig:location}).  It may be that energetic particles
did make it to 1~au, but that {\it STEREO-A}
was simply too poorly positioned in longitude to detect the event.  However, evaluating the expected intensity dependence on longitude
using the functional form and average parameters in \cite{lario13rad} (for 15-40 MeV protons), we find that the longitudinal
difference between {\it PSP} and {\it STEREO-A} magnetic footpoints would have reduced the intensity at {\it STEREO-A} by only a factor of $\sim$3, not
the factor of $\sim$20 required to make the event undetectable.  Of course, many events have been observed where intensities fall
more rapidly with radius or longitude than average \citep{lario13rad}, which evidently was the case here too.

In the event studied here, the He/H ratio of 0.052$\pm$0.002 is high compared to the average value at 1--4 MeV nucleon$^{-1}$ in large SEPs of
0.036 \citep{reames95} and compared to values in the much larger SEP events EPI-Hi observed later 
in 2019 April \citep{wiedenbeck19};
it is also much higher than the range of 0.016--0.031 found in the SIR events observed by {\it PSP} \citep{cohen19}.
However, He/H in SEPs can vary greatly from event to event and even in SIRs 
throughout the solar cycle \citep{larioheh}.  The He/H ratio in this event does
match that of 0.052$\pm$0.005 measured spectroscopically in the corona \citep{laming01}.  Although it also agrees well
with the $\sim$0.05 measured in high-speed solar wind or at solar maximum at all speeds, it is large compared to the $\sim$0.02
found in slow solar wind at solar minimum \citep{aellig01}, conditions that describe the actual observing environment.
%In EPI-Lo, the He/H ratio is strongly energy dependent over the energy range of 30--100 keV nucleon$^{-1}$ because
%the power-law spectral index for He was much harder than for H \citep{roelof19}.
Unfortunately, the small size of the event means that we have no additional composition measurements with which 
to constrain the likely source material of the energetic particles, but as described in Section~\ref{sec:obs:parts}, the
absence of heavies seems more in line with expected coronal or photospheric composition than ``typical'' SEP composition.
The extreme enhancement of heavy ions in EPI-Hi observed in the April~2 event produced by the same AR is completely
absent here only two days later (Figure~\ref{fig:zvst}). 

The differential energy spectrum averaged over the April~4 SEP event is very soft, with an index (for LET-A protons) of -4.36 on energy.
If converted to phase-space density, the index on momentum would be -10.7.   In diffusive 
shock acceleration theory, this index is equal to $-3s/(s-1)$, where $s$ is the shock compression
ratio \citep{blandford}.  So if there were a shock involved in accelerating these particles (and there is no
evidence of any), it must have been very weak, with a compression ratio of only 1.4.  The AR
was over the limb from Earth, so no measurement of any X-ray flare was possible, but as the region transited the Earth-facing 
side of the solar disk from April 6 to 19, it produced numerous small (B-class) flares.  However, compositionally the particle event does 
not exhibit the $^{3}$He and heavy ion enhancements usually found in flare-accelerated material. 
It may be that the observed plasma motion
in the EUVI surges generates sufficient compression to accelerate particles to these low energies
via some compressive acceleration mechanism \citep{roelof15,giacalone02}.  Alternatively, a ``pressure cooker''
mechanism found to accelerate ions in planetary auroras may also operate in the solar corona, and it has been
proposed as a possible SEP accelerator \citep{mitchell19}.
In this scenario, strong downward field-aligned currents confine positively charged ions that undergo perpendicular heating
by broadband electrostatic waves, gaining energy until the mirror force allows them to overcome the electric potential
and escape.  The occurrence of type III emission in this event indicates the presence of streaming electrons, a necessary requirement
for this mechanism to work.

During the April~4 event, {\it PSP} was 
0.17~au from the Sun (Figure~\ref{fig:overview}), immersed in slow solar wind with a velocity of $\sim$350 km s$^{-1}$ (Figure~\ref{fig:timeprofilesmagsw}) at 
Carrington longitude 359$^{\circ}$.  Under these conditions, the nominal Parker spiral field line connects {\it PSP} to the corona at $\sim$11$^{\circ}$
Carrington longitude.  The AR producing the EUVI surges (Figure~\ref{fig:euvi}), and
presumably the concurrent type III radio emission (Figure~\ref{fig:radio}) and the source of the April~4 SEP event, was located at 292$^{\circ}$ (Figure~\ref{fig:location}),
nearly 80$^{\circ}$ east of the nominal magnetic footpoint.  At this time, Earth was located near 60$^{\circ}$, so the 
AR was about 38$^{\circ}$ over the east limb from Earth, nearly 3~days away from making an appearance in near-Earth-based
magnetic imagers that could produce photospheric synoptic maps.  Thus, tracing the field back to the photosphere
using these maps with models such as the Current Sheet Source Surface (CSSS) model \citep{zhao95,poduval14} is prone to 
larger-than-usual uncertainties, but attempts are in progress.  Online synoptic maps from the Global Oscillation Network
Group (GONG; \url{https://gong2.nso.edu/archive/patch.pl}) of the magnetic connectivity between the
ecliptic and photosphere using the Potential Field Source Surface (PFSS; \cite{pfss}) model show conflicting
results between Carrington rotations 2215 (2019 March~12 to April~8) and 2216 (April~8 to May~5).  In rotation
2215, field lines from longitudes $\sim$290$^{\circ}$--360$^{\circ}$ map to a low-latitude extension of the 
north polar coronal hole.  The AR had not yet emerged in this synoptic map, and the field polarity
disagrees with the inward polarity measured at {\it PSP}.  By rotation 2216 the field polarity is correct, and 
field lines from the ecliptic do map back to the far western edge of AR 12738, but only over longitudes 
$\sim$300$^{\circ}$--335$^{\circ}$.  This is only half the longitudinal span required to connect to {\it PSP} at 11$^{\circ}$, which
is shown to be connected instead to the southern polar coronal hole.  

Recently, an SEP study by \cite{klassen18} reported an electron event observed by {\it STEREO} at 1 au with an impulsive rise and strong
anisotropy, clearly indicating a good magnetic connection between the flare site and the observer, 
despite the fact that the source was 90$^{\circ}$ east of the nominal magnetic footpoint.
As was the case for the event we discuss, \cite{klassen18} found that PFSS modeling did not account 
for the magnetic connection in the event they studied, at least in part because their event was also over the limb
as seen from Earth.  EUV observations
showed a very long jet propagating from the AR to the nominal magnetic footpoint of the observer, highly
inclined from the radial direction.  Although this event was much larger in 
particles and electromagnetic emission than the 2019 April~4 event and therefore better observed, the particle propagation and magnetic field
configuration may have been similar in the two events.

While the nominal magnetic footpoint of {\it PSP} was $\sim$80$^{\circ}$ from the AR, 
it was only $\sim$10$^{\circ}$ south of the heliospheric current sheet (HCS) according to 
Wilcox Solar Observatory (\url{http://wso.stanford.edu}) coronal field maps for both Carrington rotations 2215 and 2216 
(and the field polarity measured in-situ at {\it PSP} matches
that expected from these maps).  The AR was only $\sim$10$^{\circ}$ northeast of the HCS.
Perhaps solar particle transport along the HCS as described by \cite{battarbee18} played a role.
In their models, \cite{battarbee18} find that protons injected from an AR near an idealized HCS are efficiently
transported in longitude along it via drifts, westward during A$>$0 polarity solar cycles such as the one we are currently
experiencing.  The mean drift velocity along the sheet is calculated to be relatively fast, at 0.463 times the particle
speed \citep{burger85}.  Furthermore, when their modeled HCS is wavy rather than flat, escape of the particles from the HCS in
areas of large inclination (as was present locally near the AR) allows them to cross the HCS, which is required 
for particle transport during the 2019 April~4 event if the source region and HCS location have been correctly identified.

If the particle injection at the source were a step function in time, the large velocity dispersion observed (Figure~\ref{fig:vdisp})
would suggest that the average path length of the field line followed by the first-arriving particles was unusually long, a 
factor of $\sim$2 times greater than that of the nominal magnetic connection between {\it PSP} and the corona.  This seems 
physically unlikely, as the measured field was nearly radial throughout the event (Figure~\ref{fig:timeprofilesmagsw}), with
no indication of a large loop or observed CME.  Also, a step function injection would result in a sharp onset edge in a velocity dispersion plot, unlike
that observed, unless the particles underwent a significant amount of scattering, which is difficult to reconcile with the 
large anisotropy that persists throughout the event (Figure~\ref{fig:timeprofiles}).  Thus, we conclude that the particle
injection time profile was most likely not a step function, but rather extended in time.

As discussed in Section~\ref{sec:obs:parts} above, the dip in LET-A protons around 07:00 in Figure~\ref{fig:timeprofiles}
does not seem to be due to movement of the particle beam from inside to outside of LET's field of view.  There are no EUVI surges 
observed near this time (Figure~\ref{fig:euvi}) that might indicate a new injection of particles.  Both the duration and magnitude of 
the dip are similar to those of dispersionless ``dropouts'' that have been observed in small SEP events at 1~au \citep{mazur00}, in which 
field line mixing between the source and observer occasionally results in a flux tube that is
not connected to the SEP source, and thus empty of energetic particles, convecting past the observer.  However, in the case of the April~4 event, the expected dispersion
is so small that it is unclear whether or not the decrease is in fact dispersionless (Figure~\ref{fig:vdisp}).  Furthermore, the dip does not seem to be
present at the same time in LET-B, which it should be if it were the result of an empty flux tube.
As we described in Section~\ref{sec:obs:parts},
the dip does coincide with small changes in the solar wind environment. 
However, the change in energetic particle intensity (by a factor of $\sim$5) is very much larger than the change in the local
magnetic field magnitude (15\%) or solar wind density (40\%), perhaps suggesting that changes nearer the source are responsible.

%++++++++++++++++++++++++++++++

\section{Summary} \label{sec:sum}

In the {\it PSP}/{\ISOIS}/EPI-Hi instrument at energies above 1~MeV, the 2019 April~4 SEP event was soft, brief, very small, and highly anisotropic.
Specifically, the proton spectral index was -4.36 in energy, its total duration was only $\sim$15~hr, proton intensities reached the GCR background by about 8~MeV,
and event-averaged proton intensities in the generally sunward-facing aperture were $\sim$21 times greater than those in the aperture pointed 180$^{\circ}$ away,
with an additional variation of a factor of more than 10 across the sunward aperture.
No electrons or ions heavier than $^{4}$He were detected above quiet-time background levels, and no $^{3}$He nuclei were 
observed, but their intensities might be below instrument sensitivity levels in this very small event.  The absence of
heavy ions relative to He suggests that the event may be depleted in heavy elements compared with 
average SEP elemental composition.  The only
solid composition measurement possible is the He/H ratio, which at 0.052 is higher than the typical value found in SEPs \citep{reames14}, but closer
to values measured in the photosphere or corona \citep{feldman03,caffau11,lodders10}.  {\it STEREO}/EUVI images show small surges in an AR $\sim$80$^{\circ}$ east
of the nominal magnetic footpoint, temporally associated with type III emission detected by FIELDS/RFS and the EPI-Hi energetic
particles.  The large amount of particle velocity dispersion with a blurred onset edge, in the presence of a large anisotropy that persists throughout the event,
is most easily interpreted as arising from a particle injection profile that was extended in time with 
transport that was nearly scatter-free.  

No shocks or CMEs were observed in association with this event, and the very soft spectrum indicates that if any
shock were involved in accelerating the particles, it must have been very weak (with a compression ratio of $\sim$1.4).
Small flares may have been present in the AR, but were unobservable behind the limb from near-Earth X-ray detectors, 
and the particle composition does not show typical signatures of flare-accelerated material.  Acceleration involving heating
of particles confined by field-aligned currents \citep{mitchell19} is a possibility, as is compressive acceleration \citep{roelof15,giacalone02}.
It is unclear how the particles were transported 80$^{\circ}$ in longitude.  Expansion of the AR field lines between 
the photosphere and corona by this amount is not impossible, but a survey of 14 yr of PFSS maps suggests it would likely require
a region of open field lines in the photosphere at least $\sim$5$^{\circ}$ in diameter (see Figure~4 of \cite{wiedenbeck13}). 
However, as the \cite{klassen18} observations show, at times the PFSS model may simply fail to correctly identify existing highly non-radial 
field lines near ARs.
Alternatively, drift along the nearby HCS \citep{battarbee18}
may also have played a role.  Some mechanism imprinted a factor of $\sim$5 intensity drop midway through the event, which left only
a small signature in the local solar wind.

The very small size of the single near-perihelion SEP event studied here severely limits our
ability to draw definitive conclusions regarding the nature of particle acceleration and transport in it.
Conceivably, from a vantage point even closer to the source, such an event with multiple type III bursts and EUVI surges may
resolve into many discrete particle injections, just as was revealed in earlier studies comparing SEPs at 1 and 0.3~au \citep{wc06}.
The very existence of this event, during an exceptionally quiet solar minimum and completely undetectable in-situ at 1~au, 
raises the possibility that such events are more common than one might expect.  Perhaps they provide an important 
source of seed particles available for acceleration in larger SEP events (although the lack of heavy ion enrichment
in the April~4 event only 2 days after the extreme enhancement on April~2 may suggest a rather limited duration
for the presence of this seed population, at least from these very small events).  
No additional SEP events have yet been identified by
EPI-Hi inside 0.25~au, but other small near-perihelion events have 
been seen at lower energies by EPI-Lo \citep{isoisnature,roelof19,giacalone19,hill19} and at MeV energies by EPI-Hi at larger 
solar distances \citep{wiedenbeck19,schwadron19}.
Overall, the 2019 April~4 event provides a tantalizing
glimpse of what we hope to see with greater clarity as {\it PSP} gets ever closer to the Sun during its remaining 21 orbits over the
next 6 yr and solar activity picks up on the way to solar maximum.

\acknowledgments

This work was supported by NASA under contract NNN06AA01C.
{\it Parker Solar Probe} was designed, built, and is now operated by the Johns Hopkins University Applied Physics
Laboratory as part of NASA's Living With a Star (LWS) program.  Support from the LWS management and technical
team has played a critical role in the success of this mission.
We appreciate the contributions of the many individuals who have made {\it PSP} the successful mission that it is, 
in particular the EPI-Hi engineers W. R. Cook, B. Kecman, G. Dirks, and N. Angold.  We gratefully acknowledge the 
test and calibration support provided by Michigan State University's National Superconducting Cyclotron Laboratory, Texas 
A\&M University's Cyclotron Institute, and the Lawrence Berkeley National Laboratory's 88-inch Cyclotron Laboratory.
The {\ISOIS} data and visualization tools are available to the community at \url{https://spacephysics.princeton.edu/missions-instruments/isois};
data are also available via the NASA Space Physics Data Facility (\url{https://spdf.gsfc.nasa.gov/}).
We thank the GONG and Wilcox Solar Observatory teams for making their data readily accessible online.  S.D.B. acknowledges
the support of the Leverhulme Trust Visiting Professorship program.

\bibliography{apr42019}{}

\begin{thebibliography}{}
\expandafter\ifx\csname natexlab\endcsname\relax\def\natexlab#1{#1}\fi

\bibitem[{{Aellig} {et~al.}(2001){Aellig}, {Lazarus}, \&
  {Steinberg}}]{aellig01}
{Aellig}, M.~R., {Lazarus}, A.~J., \& {Steinberg}, J.~T. 2001, \grl, 28, 2767

\bibitem[{{Bale} {et~al.}(2016){Bale}, {Goetz}, {Harvey}, {Turin}, {Bonnell},
  {Dudok de Wit}, {Ergun}, {MacDowall}, {Pulupa}, {Andre}, {Bolton},
  {Bougeret}, {Bowen}, {Burgess}, {Cattell}, {Chandran}, {Chaston}, {Chen},
  {Choi}, {Connerney}, {Cranmer}, {Diaz-Aguado}, {Donakowski}, {Drake},
  {Farrell}, {Fergeau}, {Fermin}, {Fischer}, {Fox}, {Glaser}, {Goldstein},
  {Gordon}, {Hanson}, {Harris}, {Hayes}, {Hinze}, {Hollweg}, {Horbury},
  {Howard}, {Hoxie}, {Jannet}, {Karlsson}, {Kasper}, {Kellogg}, {Kien},
  {Klimchuk}, {Krasnoselskikh}, {Krucker}, {Lynch}, {Maksimovic}, {Malaspina},
  {Marker}, {Martin}, {Martinez-Oliveros}, {McCauley}, {McComas}, {McDonald},
  {Meyer-Vernet}, {Moncuquet}, {Monson}, {Mozer}, {Murphy}, {Odom},
  {Oliverson}, {Olson}, {Parker}, {Pankow}, {Phan}, {Quataert}, {Quinn},
  {Ruplin}, {Salem}, {Seitz}, {Sheppard}, {Siy}, {Stevens}, {Summers}, {Szabo},
  {Timofeeva}, {Vaivads}, {Velli}, {Yehle}, {Werthimer}, \& {Wygant}}]{fields}
{Bale}, S.~D., {Goetz}, K., {Harvey}, P.~R., {et~al.} 2016, \ssr, 204, 49

\bibitem[{{Battarbee} {et~al.}(2018){Battarbee}, {Dalla}, \&
  {Marsh}}]{battarbee18}
{Battarbee}, M., {Dalla}, S., \& {Marsh}, M.~S. 2018, \apj, 854, 23

\bibitem[{{Blandford} \& {Ostriker}(1978)}]{blandford}
{Blandford}, R.~D., \& {Ostriker}, J.~P. 1978, \apjl, 221, L29

\bibitem[{{Brueckner} {et~al.}(1995){Brueckner}, {Howard}, {Koomen},
  {Korendyke}, {Michels}, {Moses}, {Socker}, {Dere}, {Lamy}, {Llebaria},
  {Bout}, {Schwenn}, {Simnett}, {Bedford}, \& {Eyles}}]{lasco}
{Brueckner}, G.~E., {Howard}, R.~A., {Koomen}, M.~J., {et~al.} 1995, \solphys,
  162, 357

\bibitem[{{Burger} {et~al.}(1985){Burger}, {Moraal}, \& {Webb}}]{burger85}
{Burger}, R.~A., {Moraal}, H., \& {Webb}, G.~M. 1985, \apss, 116, 107

\bibitem[{{Caffau} {et~al.}(2011){Caffau}, {Ludwig}, {Steffen}, {Freytag}, \&
  {Bonifacio}}]{caffau11}
{Caffau}, E., {Ludwig}, H.-G., {Steffen}, M., {Freytag}, B., \& {Bonifacio}, P.
  2011, \solphys, 268, 255

\bibitem[{{Chae} {et~al.}(1999){Chae}, {Qiu}, {Wang}, \& {Goode}}]{chae99}
{Chae}, J., {Qiu}, J., {Wang}, H., \& {Goode}, P.~R. 1999, \apjl, 513, L75

\bibitem[{{Cohen} {et~al.}(2019){Cohen}, {Christian}, {Cummings}, {Davis},
  {Hill}, {Labrador}, {Leske}, {McComas}, {McNutt}, {Mewaldt}, {Mitchell},
  {Roelof}, {Schwadron}, {Stone}, {Szalay}, {Wiedenbeck}, {Allen}, {Ho},
  {Jian}, {Lario}, {Odstrcil}, {Bale}, {Badman}, {Pulupa}, {MacDowell},
  {Kasper}, {Case}, {Korreck}, {Larson}, \& {Stevens}}]{cohen19}
{Cohen}, C.~M.~S., {Christian}, E.~R., {Cummings}, A.~C., {et~al.} 2019, \apjs,
  in press

\bibitem[{{Domingo} {et~al.}(1995){Domingo}, {Fleck}, \& {Poland}}]{soho}
{Domingo}, V., {Fleck}, B., \& {Poland}, A.~I. 1995, \solphys, 162, 1

\bibitem[{{Feldman} \& {Widing}(2003)}]{feldman03}
{Feldman}, U., \& {Widing}, K.~G. 2003, \ssr, 107, 665

\bibitem[{{Fox} {et~al.}(2016){Fox}, {Velli}, {Bale}, {Decker}, {Driesman},
  {Howard}, {Kasper}, {Kinnison}, {Kusterer}, {Lario}, {Lockwood}, {McComas},
  {Raouafi}, \& {Szabo}}]{fox}
{Fox}, N.~J., {Velli}, M.~C., {Bale}, S.~D., {et~al.} 2016, \ssr, 204, 7

\bibitem[{{Giacalone} {et~al.}(2002){Giacalone}, {Jokipii}, \&
  {K{\'o}ta}}]{giacalone02}
{Giacalone}, J., {Jokipii}, J.~R., \& {K{\'o}ta}, J. 2002, \apj, 573, 845

\bibitem[{{Giacalone} {et~al.}(2019){Giacalone}, {Mitchell}, {Allen}, {Hill},
  {McNutt}, {Szalay}, {Desai}, {Rouillard}, {Kouloumvakos}, {McComas},
  {Christian}, {Schwadron}, {Bale}, {Case}, {Chen}, {Cohen}, {Joyce}, {Kasper},
  {Klein}, {Korreck}, {Krimigis}, {Larson}, {Livi}, {Leske}, {MacDowell},
  {Matthaeus}, {Mewaldt}, {Nieves-Chinchilla}, {Pulupa}, {Roelof}, {Stevens},
  {Stone}, {Szabo}, \& {Whittlesey}}]{giacalone19}
{Giacalone}, J., {Mitchell}, D.~G., {Allen}, R.~C., {et~al.} 2019, \apjs, in
  press

\bibitem[{{Hill} {et~al.}(2017){Hill}, {Mitchell}, {Andrews}, {Cooper},
  {Gurnee}, {Hayes}, {Layman}, {McNutt}, {Nelson}, {Parker}, {Schlemm},
  {Stokes}, {Begley}, {Boyle}, {Burgum}, {Do}, {Dupont}, {Gold}, {Haggerty},
  {Hoffer}, {Hutcheson}, {Jaskulek}, {Krimigis}, {Liang}, {London}, {Noble},
  {Roelof}, {Seifert}, {Strohbehn}, {Vandegriff}, \& {Westlake}}]{mushroom}
{Hill}, M.~E., {Mitchell}, D.~G., {Andrews}, G.~B., {et~al.} 2017, Journal of
  Geophysical Research (Space Physics), 122, 1513

\bibitem[{{Hill} {et~al.}(2019){Hill}, {Mitchell}, {Allen}, {Brown}, {de
  Nolfo}, {McNutt}, {Mitchell}, {Szalay}, {Vourlidas}, {Arge}, {Christian},
  {Cohen}, {Crew}, {Desai}, {Giacalone}, {Henney}, {Jones}, {Joyce},
  {Krimigis}, {Leske}, {McComas}, {Mewaldt}, {Nelson}, {Niehof}, {Roelof},
  {Schwadron}, {Wallace}, \& {Wiedenbeck}}]{hill19}
{Hill}, M.~E., {Mitchell}, D.~G., {Allen}, R.~C., {et~al.} 2019, \apjs, in
  press

\bibitem[{{Howard} {et~al.}(2008){Howard}, {Moses}, {Vourlidas}, {Newmark},
  {Socker}, {Plunkett}, {Korendyke}, {Cook}, {Hurley}, {Davila}, {Thompson},
  {St Cyr}, {Mentzell}, {Mehalick}, {Lemen}, {Wuelser}, {Duncan}, {Tarbell},
  {Wolfson}, {Moore}, {Harrison}, {Waltham}, {Lang}, {Davis}, {Eyles},
  {Mapson-Menard}, {Simnett}, {Halain}, {Defise}, {Mazy}, {Rochus}, {Mercier},
  {Ravet}, {Delmotte}, {Auchere}, {Delaboudiniere}, {Bothmer}, {Deutsch},
  {Wang}, {Rich}, {Cooper}, {Stephens}, {Maahs}, {Baugh}, {McMullin}, \&
  {Carter}}]{secchi}
{Howard}, R.~A., {Moses}, J.~D., {Vourlidas}, A., {et~al.} 2008, \ssr, 136, 67

\bibitem[{{Innes} {et~al.}(2009){Innes}, {Genetelli}, {Attie}, \&
  {Potts}}]{innes}
{Innes}, D.~E., {Genetelli}, A., {Attie}, R., \& {Potts}, H.~E. 2009, \aap,
  495, 319

\bibitem[{{Joyce} {et~al.}(2019){Joyce}, {McComas}, {Christian}, {Schwadron},
  {Wiedenbeck}, {McNutt}, {Cohen}, {Leske}, {Mewaldt}, {Stone}, {Labrador},
  {Davis}, {Cummings}, {Mitchell}, {Hill}, {Roelof}, {Szalay}, {Rankin},
  {Desai}, {Giacalone}, \& {Matthaeus}}]{joyce19}
{Joyce}, C.~J., {McComas}, D.~J., {Christian}, E.~R., {et~al.} 2019, \apjs, in
  press

\bibitem[{{Kaiser} {et~al.}(2008){Kaiser}, {Kucera}, {Davila}, {St.~Cyr},
  {Guhathakurta}, \& {Christian}}]{stereo}
{Kaiser}, M.~L., {Kucera}, T.~A., {Davila}, J.~M., {et~al.} 2008, \ssr, 136, 5

\bibitem[{{Kasper} {et~al.}(2016){Kasper}, {Abiad}, {Austin}, {Balat-Pichelin},
  {Bale}, {Belcher}, {Berg}, {Bergner}, {Berthomier}, {Bookbinder}, {Brodu},
  {Caldwell}, {Case}, {Chandran}, {Cheimets}, {Cirtain}, {Cranmer}, {Curtis},
  {Daigneau}, {Dalton}, {Dasgupta}, {DeTomaso}, {Diaz-Aguado}, {Djordjevic},
  {Donaskowski}, {Effinger}, {Florinski}, {Fox}, {Freeman}, {Gallagher},
  {Gary}, {Gauron}, {Gates}, {Goldstein}, {Golub}, {Gordon}, {Gurnee}, {Guth},
  {Halekas}, {Hatch}, {Heerikuisen}, {Ho}, {Hu}, {Johnson}, {Jordan},
  {Korreck}, {Larson}, {Lazarus}, {Li}, {Livi}, {Ludlam}, {Maksimovic},
  {McFadden}, {Marchant}, {Maruca}, {McComas}, {Messina}, {Mercer}, {Park},
  {Peddie}, {Pogorelov}, {Reinhart}, {Richardson}, {Robinson}, {Rosen},
  {Skoug}, {Slagle}, {Steinberg}, {Stevens}, {Szabo}, {Taylor}, {Tiu}, {Turin},
  {Velli}, {Webb}, {Whittlesey}, {Wright}, {Wu}, \& {Zank}}]{sweap}
{Kasper}, J.~C., {Abiad}, R., {Austin}, G., {et~al.} 2016, \ssr, 204, 131

\bibitem[{{Klassen} {et~al.}(2018){Klassen}, {Dresing}, {G{\'o}mez-Herrero},
  {Heber}, \& {Veronig}}]{klassen18}
{Klassen}, A., {Dresing}, N., {G{\'o}mez-Herrero}, R., {Heber}, B., \&
  {Veronig}, A. 2018, \aap, 614, A61

\bibitem[{{Laming} \& {Feldman}(2001)}]{laming01}
{Laming}, J.~M., \& {Feldman}, U. 2001, \apj, 546, 552

\bibitem[{{Lario} {et~al.}(2013){Lario}, {Aran}, {G{\'o}mez-Herrero},
  {Dresing}, {Heber}, {Ho}, {Decker}, \& {Roelof}}]{lario13rad}
{Lario}, D., {Aran}, A., {G{\'o}mez-Herrero}, R., {et~al.} 2013, \apj, 767, 41

\bibitem[{{Lario} {et~al.}(2003){Lario}, {Roelof}, {Decker}, {Ho}, {Maclennan},
  \& {Gosling}}]{larioheh}
{Lario}, D., {Roelof}, E.~C., {Decker}, R.~B., {et~al.} 2003, Annales
  Geophysicae, 21, 1229

\bibitem[{{Lodders}(2010)}]{lodders10}
{Lodders}, K. 2010, Astrophysics and Space Science Proceedings, 16, 379

\bibitem[{{Mazur} {et~al.}(2000){Mazur}, {Mason}, {Dwyer}, {Giacalone},
  {Jokipii}, \& {Stone}}]{mazur00}
{Mazur}, J.~E., {Mason}, G.~M., {Dwyer}, J.~R., {et~al.} 2000, \apjl, 532, L79

\bibitem[{{McComas} {et~al.}(2016){McComas}, {Alexander}, {Angold}, {Bale},
  {Beebe}, {Birdwell}, {Boyle}, {Burgum}, {Burnham}, {Christian}, {Cook},
  {Cooper}, {Cummings}, {Davis}, {Desai}, {Dickinson}, {Dirks}, {Do}, {Fox},
  {Giacalone}, {Gold}, {Gurnee}, {Hayes}, {Hill}, {Kasper}, {Kecman}, {Klemic},
  {Krimigis}, {Labrador}, {Layman}, {Leske}, {Livi}, {Matthaeus}, {McNutt},
  {Mewaldt}, {Mitchell}, {Nelson}, {Parker}, {Rankin}, {Roelof}, {Schwadron},
  {Seifert}, {Shuman}, {Stokes}, {Stone}, {Vandegriff}, {Velli}, {von
  Rosenvinge}, {Weidner}, {Wiedenbeck}, \& {Wilson}}]{isois}
{McComas}, D.~J., {Alexander}, N., {Angold}, N., {et~al.} 2016, Space Science
  Reviews, 204, 187

\bibitem[{{McComas} {et~al.}(2019){McComas}, {Christian}, {Cohen}, {Cummings},
  {Davis}, {Desai}, {Giacalone}, {Hill}, {Joyce}, {Krimigis}, {Labrador},
  {Leske}, {Malandraki}, {Matthaeus}, {McNutt}, {Mewaldt}, {Mitchell},
  {Posner}, {Rankin}, {Roelof}, {Schwadron}, {Stone}, {Szalay}, {Wiedenbeck},
  {Bale}, {Kasper}, {Case}, {Korreck}, {MacDowall}, {Pulupa}, {Stevens}, \&
  {Rouillard}}]{isoisnature}
{McComas}, D.~J., {Christian}, E.~R., {Cohen}, C.~M.~S., {et~al.} 2019, Nature,
  576, 223

\bibitem[{{Mewaldt} {et~al.}(2008){Mewaldt}, {Cohen}, {Cook}, {Cummings},
  {Davis}, {Geier}, {Kecman}, {Klemic}, {Labrador}, {Leske}, {Miyasaka},
  {Nguyen}, {Ogliore}, {Stone}, {Radocinski}, {Wiedenbeck}, {Hawk}, {Shuman},
  {von Rosenvinge}, \& {Wortman}}]{stereolet}
{Mewaldt}, R.~A., {Cohen}, C.~M.~S., {Cook}, W.~R., {et~al.} 2008, \ssr, 136,
  285

\bibitem[{{Mitchell} {et~al.}(2019){Mitchell}, {Giacalone}, {Allen}, {Hill},
  {McNutt}, {McComas}, {Szalay}, {Schwadron}, {Rouillard}, {Bale}, {Pulupa},
  {Kasper}, {MacDowell}, {Christian}, {Wiedenbeck}, \&
  {Matthaeus}}]{mitchell19}
{Mitchell}, D.~G., {Giacalone}, J., {Allen}, R.~C., {et~al.} 2019, \apj,
  submitted

\bibitem[{{Podladchikova} {et~al.}(2010){Podladchikova}, {Vourlidas}, {Van der
  Linden}, {W{\"u}lser}, \& {Patsourakos}}]{pod2010}
{Podladchikova}, O., {Vourlidas}, A., {Van der Linden}, R.~A.~M., {W{\"u}lser},
  J.~P., \& {Patsourakos}, S. 2010, \apj, 709, 369

\bibitem[{{Poduval} \& {Zhao}(2014)}]{poduval14}
{Poduval}, B., \& {Zhao}, X.~P. 2014, \apjl, 782, L22

\bibitem[{{Pulupa} {et~al.}(2017){Pulupa}, {Bale}, {Bonnell}, {Bowen},
  {Carruth}, {Goetz}, {Gordon}, {Harvey}, {Maksimovic},
  {Mart{\'\i}nez-Oliveros}, {Moncuquet}, {Saint-Hilaire}, {Seitz}, \&
  {Sundkvist}}]{pulupa}
{Pulupa}, M., {Bale}, S.~D., {Bonnell}, J.~W., {et~al.} 2017, Journal of
  Geophysical Research (Space Physics), 122, 2836

\bibitem[{{Reames}(1995)}]{reames95}
{Reames}, D.~V. 1995, Advances in Space Research, 15, (7)41

\bibitem[{{Reames}(2014)}]{reames14}
---. 2014, \solphys, 289, 977

\bibitem[{{Roelof}(2015)}]{roelof15}
{Roelof}, E.~C. 2015, J. Phys.: Conf. Ser., 642, 012023

\bibitem[{{Roelof} {et~al.}(2019){Roelof}, {Allen}, {Bale}, {Christian},
  {Cohen}, {Cummings}, {Hill}, {Leske}, {McComas}, {McNutt}, {Mewaldt},
  {Mitchell}, {Pulupa}, {Schwadron}, {Stone}, {Szalay}, {Wiedenbeck}, \&
  {Vourlidas}}]{roelof19}
{Roelof}, E.~C., {Allen}, R.~C., {Bale}, S.~D., {et~al.} 2019, \apj, in
  preparation

\bibitem[{{Roy}(1973)}]{roy73}
{Roy}, J.-R. 1973, \solphys, 32, 139

\bibitem[{{Schmahl}(1981)}]{schmahl81}
{Schmahl}, E.~J. 1981, \solphys, 69, 135

\bibitem[{{Schwadron} {et~al.}(2019){Schwadron}, {Bale}, {Bonnell}, {Case},
  {Christian}, {Cohen}, {Cummings}, {Davis}, {Dudok de Wit}, {de Wet}, {Desai},
  {Joyce}, {Goetz}, {Giacalone}, {Gorby}, {Harvey}, {Heber}, {Hill},
  {Karavolos}, {Kasper}, {Korreck}, {Krimigis}, {Larson}, {Livi}, {Leske},
  {Malandraki}, {MacDowell}, {Malaspina}, {Matthaeus}, {McComas}, {McNutt},
  {Mewaldt}, {Mitchell}, {Mays}, {Niehof}, {Odstrcil}, {Pulupa}, {Poduval},
  {Posner}, {Rankin}, {Roelof}, {Stevens}, {Stone}, {Szalay}, {Wiedenbeck},
  {Winslow}, \& {Whittlesey}}]{schwadron19}
{Schwadron}, N.~A., {Bale}, S., {Bonnell}, J., {et~al.} 2019, \apjs, in press

\bibitem[{{Stenborg} {et~al.}(2008){Stenborg}, {Vourlidas}, \&
  {Howard}}]{stenborg08}
{Stenborg}, G., {Vourlidas}, A., \& {Howard}, R.~A. 2008, \apj, 674, 1201

\bibitem[{{Vourlidas} {et~al.}(2003){Vourlidas}, {Wu}, {Wang}, {Subramanian},
  \& {Howard}}]{vourlidas03}
{Vourlidas}, A., {Wu}, S.~T., {Wang}, A.~H., {Subramanian}, P., \& {Howard},
  R.~A. 2003, \apj, 598, 1392

\bibitem[{{Wang} \& {Sheeley}(1992)}]{pfss}
{Wang}, Y.-M., \& {Sheeley}, Jr., N.~R. 1992, \apj, 392, 310

\bibitem[{{Wibberenz} \& {Cane}(2006)}]{wc06}
{Wibberenz}, G., \& {Cane}, H.~V. 2006, \apj, 650, 1199

\bibitem[{{Wiedenbeck} {et~al.}(2013){Wiedenbeck}, {Mason}, {Cohen}, {Nitta},
  {G{\'o}mez-Herrero}, \& {Haggerty}}]{wiedenbeck13}
{Wiedenbeck}, M.~E., {Mason}, G.~M., {Cohen}, C.~M.~S., {et~al.} 2013, \apj,
  762, 54

\bibitem[{{Wiedenbeck} {et~al.}(2017){Wiedenbeck}, {Angold}, {Birdwell},
  {Burnham}, {Christian}, {Cohen}, {Cook}, {Cummings}, {Davis}, {Dirks}, {Do},
  {Everett}, {Goodwin}, {Hanley}, {Hernandez}, {Kecman}, {Klemic}, {Labrador},
  {Leske}, {Lopez}, {Link}, {McComas}, {Mewaldt}, {Miyasaka}, {Nahory},
  {Rankin}, {Riggans}, {Rodriguez}, {Rusert}, {Shuman}, {Simms}, {Stone}, {von
  Rosenvinge}, {Weidner}, \& {White}}]{wiedenbeck2017a}
{Wiedenbeck}, M.~E., {Angold}, N.~G., {Birdwell}, B., {et~al.} 2017, in
  International Cosmic Ray Conference, Vol. 301, 35th International Cosmic Ray
  Conference (ICRC2017), 16

\bibitem[{{Wiedenbeck} {et~al.}(2019){Wiedenbeck}, {Allen}, {Bale}, R.,
  {Christian}, {Cohen}, {Cummings}, {Davis}, {Desai}, {Giacalone}, {Hill},
  {Ho}, {Joyce}, {Kasper}, {Krimigis}, {Leske}, {Labrador}, {Malandraki},
  {Mason}, {Matthaeus}, {McComas}, {McNutt}, {Mewaldt}, {Mitchell}, {Posner},
  {Pulupa}, {Rankin}, {Roelof}, {Schwadron}, {Stone}, \&
  {Szalay}}]{wiedenbeck19}
{Wiedenbeck}, M.~E., {Allen}, R.~C., {Bale}, S., {et~al.} 2019, \apjs, in press

\bibitem[{{Zhao} \& {Hoeksema}(1995)}]{zhao95}
{Zhao}, X., \& {Hoeksema}, J.~T. 1995, \jgr, 100, 19

\end{thebibliography}
\bibliographystyle{aasjournal}

\listofchanges
\end{document}